\documentclass[aps,prl,twocolumn,showpacs,showkeys,tightenlines,superscriptaddress,longbibliography]{revtex4-2}

\usepackage{amsmath,amstext,amsfonts,amsbsy,amssymb,amscd,slashed}
\usepackage{textcomp}
\usepackage{color,svg,epsfig,hyperref,bbm,multirow}
\usepackage{graphicx}
\usepackage{tabularx}
\usepackage{booktabs}
\usepackage{floatpag}
\usepackage{physics}
\usepackage{hyperref}

\usepackage[makeroom]{cancel}
\usepackage{blkarray}
\usepackage{stackengine}

\usepackage{tikz}
\usetikzlibrary{arrows,shapes}
\usetikzlibrary{trees}
\usetikzlibrary{matrix,arrows}                          
\usetikzlibrary{positioning}                            
\usetikzlibrary{calc,through}                           
\usetikzlibrary{decorations.pathreplacing}  
\usetikzlibrary{decorations.pathmorphing}       
\usetikzlibrary{decorations.markings}
\tikzset{
    vector/.style={decorate, decoration={snake}, draw},
        provector/.style={decorate, decoration={snake,amplitude=2.5pt}, draw},
        antivector/.style={decorate, decoration={snake,amplitude=-2.5pt}, draw},
    fermion/.style={draw=black, postaction={decorate},
        decoration={markings,mark=at position .55 with {\arrow[draw=black]{>}}}},
    fermionbar/.style={draw=black, postaction={decorate},
        decoration={markings,mark=at position .55 with {\arrow[draw=black]{<}}}},
    fermionnoarrow/.style={draw=black},
    gluon/.style={decorate, draw=black,
        decoration={coil,amplitude=4pt, segment length=5pt}},                           
    scalar/.style={dashed,draw=black, postaction={decorate},
        decoration={markings,mark=at position .55 with {\arrow[draw=black]{>}}}},
    scalarbar/.style={dashed,draw=black, postaction={decorate},
        decoration={markings,mark=at position .55 with {\arrow[draw=black]{<}}}},
    scalarnoarrow/.style={dashed,draw=black},
    electron/.style={draw=black, postaction={decorate},
        decoration={markings,mark=at position .55 with {\arrow[draw=black]{>}}}},
        bigvector/.style={decorate, decoration={snake,amplitude=4pt}, draw},
}

\newcolumntype{C}{>{$}c<{$}}
\newcolumntype{R}{>{$}r<{$}}
\newcolumntype{L}{>{$}l<{$}}

\usepackage[normalem]{ulem}
\newcommand\xoutpars[1]{\let\helpcmd\xout\parhelp#1\par\relax\relax}
\newcommand\soutpars[1]{\let\helpcmd\sout\parhelp#1\par\relax\relax}
\long\def\parhelp#1\par#2\relax{%
  \helpcmd{#1}\ifx\relax#2\else\par\parhelp#2\relax\fi%
}

\begin{document}
	\vspace*{-30truemm}
	\begin{flushright}
		MITP-26-001
		\\
		CERN-TH-2026-004
        \\
        KEK-TH-2795
	\end{flushright}

	
\title{Running of the Electroweak Gauge Couplings from First Principles}

\author{Alessandro~Conigli}
\email{aconigli@uni-mainz.de}
\affiliation{Helmholtz Institute Mainz, Johannes Gutenberg-Universit\"{a}t
Mainz, 55099 Mainz, Germany}
\affiliation{GSI Helmholtz Centre for Heavy Ion Research, 64291 Darmstadt, Germany
}
\author{Dalibor~Djukanovic}
\affiliation{Helmholtz Institute Mainz, Johannes Gutenberg-Universit\"{a}t
Mainz, 55099 Mainz, Germany}
\affiliation{GSI Helmholtz Centre for Heavy Ion Research, 64291 Darmstadt, Germany
}
\author{Georg~von~Hippel}
\affiliation{
 PRISMA$^{++}$ Cluster of Excellence and Institut f\"{u}r Kernphysik, Johannes Gutenberg-Universit\"{a}t Mainz, 55099 Mainz, Germany
}
\author{Simon~Kuberski}
\affiliation{
Theoretical Physics Department, CERN, 1211 Geneva 23, Switzerland
}
\author{Harvey~B.~Meyer}
\affiliation{Helmholtz Institute Mainz, Johannes Gutenberg-Universit\"{a}t
Mainz, 55099 Mainz, Germany}
\affiliation{
 PRISMA$^{++}$ Cluster of Excellence and Institut f\"{u}r Kernphysik, Johannes Gutenberg-Universit\"{a}t Mainz, 55099 Mainz, Germany
}
\affiliation{
Theoretical Physics Department, CERN, 1211 Geneva 23, Switzerland
}
\author{Kohtaroh~Miura}
\affiliation{ Institute of Particle and Nuclear Studies, High Energy Accelerator Research Organization (KEK), Tsukuba 305-0801, Japan
}
\author{Konstantin~Ottnad}
\altaffiliation[Current address: ]{Helmholtz-Institut f\"{u}r Strahlen- und Kernphysik and Bethe Center for Theoretical Physics, Universit\"{a}t Bonn, D-53115 Bonn, Germany}
\affiliation{
 PRISMA$^{++}$ Cluster of Excellence and Institut f\"{u}r Kernphysik, Johannes Gutenberg-Universit\"{a}t Mainz, 55099 Mainz, Germany
}
\author{Andreas~Risch}
\affiliation{Department of Physics, University of Wuppertal, Gaussstr. 20, 42119
Wuppertal, Germany}

\author{Hartmut~Wittig}
\affiliation{Helmholtz Institute Mainz, Johannes Gutenberg-Universit\"{a}t
Mainz, 55099 Mainz, Germany}
\affiliation{GSI Helmholtz Centre for Heavy Ion Research, 64291 Darmstadt, Germany
}
\affiliation{
 PRISMA$^{++}$ Cluster of Excellence and Institut f\"{u}r Kernphysik, Johannes Gutenberg-Universit\"{a}t Mainz, 55099 Mainz, Germany
}

\begin{abstract}
	We present a high-precision calculation of the hadronic running of electroweak gauge couplings from first principles. Employing lattice QCD in the low-energy regime, we achieve permille precision for virtualities $Q^2 \lesssim 12\;\mathrm{GeV}^2$. 
	At $Q^2 \simeq 1\;\mathrm{GeV}^2$, our  determination deviates by up to $7\sigma$ from estimates based on $e^+e^-$ measurements. Combining  lattice QCD  with perturbative QCD via the Euclidean split technique, we obtain for the electromagnetic coupling
	$\Delta\alpha^{(5)}_{\mathrm{had}}(M_Z^2) = 0.027821(34)_{\mathrm{lat}}(35)_{\mathrm{pQCD}}$, which is more than twice as precise as recent phenomenological determinations. 
	We assess improvement scenarios by which the precision target for next-generation electroweak measurements could be reached.
\end{abstract}

\maketitle


\paragraph{Introduction}
The electromagnetic coupling and the weak mixing angle are fundamental parameters of the Standard Model (SM). They play a pivotal role in the global effort to detect signals of beyond-Standard Model (BSM) physics. In particular, the value of the electromagnetic coupling at the $Z$-pole, $\alpha(M_Z^2)^{-1}=127.930\pm0.008$~\cite{ParticleDataGroup:2024cfk} is an important input parameter for interpreting the results from high-energy colliders. The uncertainty on $\alpha(M_Z^2)$ is dominated by the hadronic contribution to its running, $\Delta\alpha_{\mathrm{had}}^{(5)}$, which therefore represents a key quantity for stringent tests of the SM. The same Hadronic Vacuum Polarization (HVP) also enters the leading hadronic contribution to the muon anomalous magnetic moment, $a_\mu^{\mathrm{hvp}}$, providing a nontrivial connection between low- and high-energy precision observables.
 Improving $\Delta\alpha_{\mathrm{had}}^{(5)}$ directly enhances the sensitivity of future electroweak measurements. The FCC-ee aims at an absolute precision of $(3-5)\times 10^{-5}$  on $\alpha(M_Z^2)$, based on measurements of the muon forward-backward asymmetry~\cite{Janot:2015gjr,Freitas:2019bre,Jeger_yellow_rep}. This corresponds to a required $(1.0-1.7)$ \textperthousand $\,$ precision on the hadronic contribution, with the upper bound consistent with the uncertainty achieved in this Letter. In addition, a recent proposal based on $Z$-pole measurements has been shown to reach a projected statistical sensitivity below the $10^{-5}$ level \cite{Riembau:2025ppc}.

A complementary test of the SM is provided by the scale dependence of the weak mixing angle, $\sin^2\theta_W$. Upcoming low-$Q^2$ parity-violating experiments, such as P2 $(e‑p)$~\cite{Becker:2018ggl} and MOLLER $(e‑e)$~\cite{MOLLER:2014iki}, will provide benchmarks with competitive precision.

The hadronic contributions constitute the leading source of uncertainty in the SM predictions for $\Delta\alpha_{\mathrm{had}}^{(5)}$ and $\sin^2\theta_W$. Traditional dispersive evaluations rely on $e^+e^-\to\hbox{hadrons}$   cross sections data~\cite{Keshavarzi:2018mgv, Davier:2019can,Jeger_yellow_rep, Jegerlehner:2019alphaQEDc19}, which are sensitive to unresolved tensions in the $\pi^+\pi^-$ channel. This issue motivated the Muon $g-2$ Theory Initiative \cite{Aliberti:2025beg} to omit the data-driven $a_\mu^{\text{hvp}}$ estimates, favoring a fully first-principles approach via lattice QCD. For $\sin^2\theta_W$, the data-driven approach is further complicated by model-dependent flavor separation.

Here we report a new lattice calculation with significantly reduced uncertainties compared to our previous work~\cite{Ce:2022eix}. Our final result for the energy dependence (running)
of $\alpha$ at the $Z$-pole is
\begin{equation}
	\Delta\alpha_{\mathrm{had}}^{(5)}(M_Z^2) =
	0.027\,821(34)_{\mathrm{lat}}(35)_{\mathrm{pQCD}}\,,
\end{equation}
achieving a total relative uncertainty of 0.17\%, and improving upon state-of-the-art phenomenological determinations ~\cite{Davier:2017zfy, Keshavarzi:2018mgv, Davier:2019can, Keshavarzi:2019abf, Jeger_yellow_rep, Jegerlehner:2019alphaQEDc19} by about a factor of 2.

\paragraph{Methodology}

The running of the electromagnetic coupling $\alpha(q^2)$
and the electroweak mixing angle $\sin^2\theta_W(q^2)$ can be calculated using
renormalization group techniques.  In the on-shell scheme, the running of
$\alpha$ is expressed as \begin{equation} \alpha(q^2) = \frac{\alpha}{1 -
\Delta\alpha(q^2)}, \end{equation} where $\alpha^{-1} = 137.035999178(8)$ is the
Thomson-limit value. The leptonic contribution to $\Delta\alpha$ is known to high precision from
perturbative QED, but the hadronic part 
\begin{align} 
\Delta\alpha_{\rm
had}(q^2) &= 4\pi\alpha\,
\Re\big[\bar\Pi(q^2)\big],\nonumber\\
 \,\bar\Pi(q^2) &=
\Pi(q^2) - \Pi(0), 
\end{align} 
where $\bar\Pi(q^2)$ denotes the subtracted HVP\@, must be determined non-perturbatively.
Traditionally, this is done via a dispersion relation using the experimentally
measured $R$ ratio~\cite{Erler:2017knj, Proceedings:2019vxr, Davier:2019can,
Keshavarzi:2019abf}, yielding the world average $\Delta\alpha_{\rm
had}^{(5)}(M_Z^2) = 0.02783(6)$~\cite{ParticleDataGroup:2024cfk}. Lattice QCD
provides an alternative, first-principles approach by computing the Euclidean
correlator of the electromagnetic current
$j_\mu^\gamma$~\cite{Burger:2015lqa,Francis:2015grz,
Budapest-Marseille-Wuppertal:2017okr}, which directly encodes the HVP
$\bar{\Pi}^{(\gamma,\gamma)}(q^2)$.

Similarly, the running of $\sin^2\theta_W$ is expressed in terms of the weak-isospin coupling $\alpha_2(q^2) = \alpha_2 / (1 -
\Delta\alpha_2(q^2))$~\cite{Jeger1986, Jeger_yellow_rep, Jegerlehner:2011mw,
	Jegerlehner:2017zsb}, for which the hadronic contribution is  
\begin{align} 
(\Delta\sin^2\theta_W)_{\rm had}(q^2)
&= \Delta\alpha_{\rm had}(q^2) - \Delta\alpha_{2,\rm had}(q^2) \nonumber\\
& =
-\frac{4\pi\alpha}{\sin^2\theta_W(0)}\bar\Pi^{(Z,\gamma)}(q^2),
\end{align} 
where $\bar\Pi^{(Z,\gamma)}$ denotes the mixing between the
electromagnetic current and the vector part of the neutral weak current
$j_\mu^Z$. We adopt the value $\sin^2\theta_W(0)=0.23857(5)$ from Ref.~\cite{ParticleDataGroup:2020ssz}, obtained from the average of several determinations~\cite{Czarnecki:2000ic, Kumar:2013yoa, Erler:2004in, Erler:2017knj}.

To compute these quantities, we use the time-momentum representation
(TMR)~\cite{Bernecker:2011gh, Francis:2013fzp},
which relates the subtracted HVP  to the Euclidean vector correlator
\begin{equation} 
G^{(\alpha,\gamma)}(x_0) = -\frac{1}{3}\int \dd^3x\,\big\langle
j^\alpha_k(x_0,\mathbf{x})\, j^\gamma_k(0)\big\rangle, \,\, \alpha =
\gamma,\,Z, 
\end{equation} 
via the integral\footnote{Here we adopt the convention  $Q^2=-q^2$ for space-like Euclidean momenta.}

\begin{equation}
	\bar\Pi^{(\alpha,\gamma)}(-Q^2) = \int_0^\infty
	\dd x_0\,G^{(\alpha,\gamma)}(x_0)\, K(x_0, Q^2),
\end{equation} 
where $K(x_0, Q^2)$ is a $Q^2$-dependent kernel function.

To improve chiral-continuum extrapolations and isolate regions sensitive to different Euclidean distances, we use a telescopic window decomposition inspired by the short, intermediate and long distance windows introduced in ~\cite{RBC:2018dos}.
The contributions to the HVP are split into three domains, via
\begin{align}   \label{eq:hvp_splitting}
            \bar{\Pi}(-Q^2)
          &= \widehat{\Pi}(-Q^2) + \widehat{\Pi}(-Q^2/4) + \bar{\Pi}(-Q^2/16),
  \end{align}
where 
\begin{equation}
     \widehat{\Pi}(-Q^2) \equiv     \Pi(-Q^2) - \Pi(-Q^2/4).
    \end{equation}
This defines three kinematic regions: the high-virtuality (HV) contribution, $\widehat{\Pi}(-Q^2)$,
dominated by short-distance physics and most sensitive to discretization effects; the mid-virtuality (MV) piece, $ \widehat{\Pi}(-Q^2/4)$, arising from intermediate distances;  and  the low-virtuality (LV) component, $ \bar{\Pi}(-Q^2/16)$, which receives the dominant long-distance contribution and benefits most from enhanced statistics and noise-reduction techniques.

Our analysis uses 27 CLS ensembles with lattice spacings
$a\in[0.039,0.085]\,\mathrm{fm}$ and pion masses down to the physical
point~\cite{Bruno:2014jqa,Bali:2016umi, Mohler:2017wnb, Mohler:2020txx,
Kuberski:2023zky}. Noise reduction in the long-distance Euclidean region is achieved via low-mode
averaging~\cite{Giusti:2004yp,DeGrand:2004qw} and spectral reconstruction as
applied in~\cite{Djukanovic:2024cmq}. Finite-volume effects are included via 
a hybrid Hansen-Patella~\cite{Hansen:2019rbh, Hansen:2020whp} and Meyer-Lellouch-L\"uscher~\cite{Lellouch:2000pv, Meyer:2011um} framework. We first match all ensembles to a common reference volume,  perform chiral-continuum extrapolations, and  apply the final infinite volume correction in the continuum following  Ref.~\cite{Djukanovic:2024cmq}. 
The extrapolation to physical pion mass and vanishing lattice spacing is performed through  a global fit of the quark mass and lattice spacing dependence, following Refs.~\cite{Kuberski:2024bcj, Djukanovic:2024cmq, companion_paper}. 
The continuum limit is obtained via a controlled  lattice spacing extrapolation, and the physical scale is set using Ref.~\cite{Bussone:2025wlf}.  Systematic uncertainties are estimated by varying the fit ansatz and combining the results using a model averaging procedure.  In addition to variations of the fit forms we also apply cuts on the ensemble set entering each fit to assess systematic effects. Final central values and systematic uncertainties are obtained  using the methods outlined in~\cite{Jay:2020jkz}, with weights based on the Akaike
Information Criterion (AIC)~\cite{Akaike:1998zah}.
 Chiral-continuum extrapolations are performed separately for the isovector, isoscalar, and charm-connected channels and for each  virtuality region, as the most suitable functional forms depend on both isospin and virtuality. A full description of the numerical setup, noise reduction strategies and fit variants is provided in the companion paper~\cite{companion_paper}. The chiral-continuum extrapolation plots and fit quality indicators for all channels entering the analysis are collected in the Supplemental Material~\cite{SuppMat}. 

\begin{figure}
	\centering
	\includegraphics[scale=0.255]{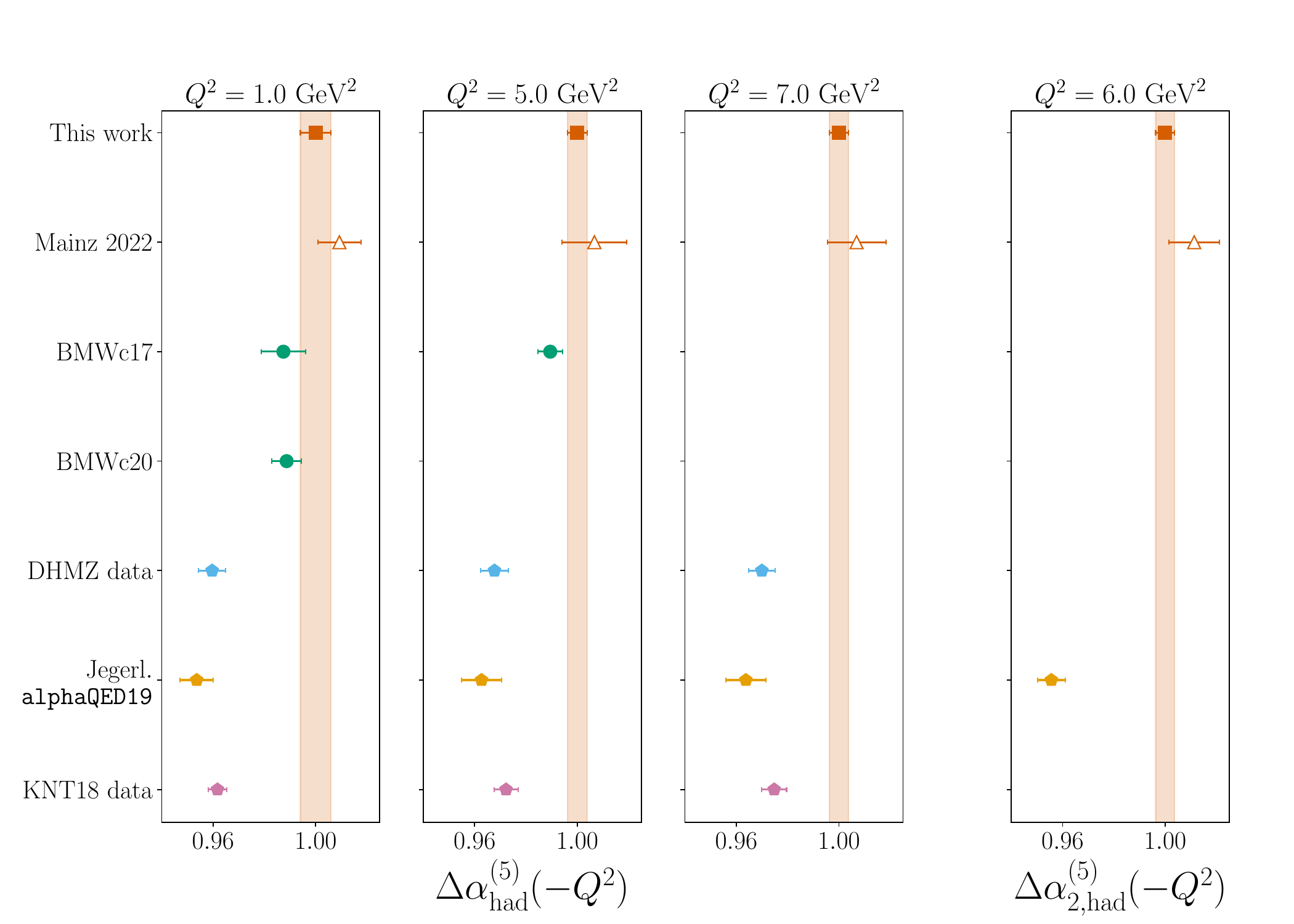}
	\caption{Summary of results for $\Delta\alpha_{\mathrm{had}}^{(5)}(-Q^2)$ from lattice and phenomenological determinations shown in the first three panels, and for $\Delta\alpha_{2,\mathrm{had}}^{(5)}(-Q^2)$ in the last panel. Shown are ratios  to our central  values, highlighting relative deviations. For our calculation, the orange vertical  band denotes the total uncertainty, including the small bottom-quark contribution. Subleading isospin-breaking effects are not included.}
	\label{fig:comparison_space_like_region} 
\end{figure}

We estimate isospin-breaking corrections based on results obtained
for the quark-connected contributions to $\bar \Pi$ following the approach of ~\cite{Ce:2022eix, Ce:2022kxy, Kuberski:2024bcj, Djukanovic:2024cmq}, supplemented by an independent phenomenological estimate based on the model of~\cite{Biloshytskyi:2022ets} with inputs from~\cite{Parrino:2025afq,Erb:2025nxk}. 
Combining both approaches, we find a small isospin-breaking correction at $Q^2=9 \ \mathrm{GeV}^2$,  
        \begin{eqnarray}\label{eq:ib_effects}
                  \Delta_{\rm IB} \Delta\alpha_{\mathrm{had}}(-Q^2) &=& (0.75\pm1.28) \times
10^{-5},
\\ \nonumber
 \Delta_{\rm IB} (\Delta\sin^2\theta_W)_{\mathrm{had}}(-Q^2) &=& (-0.82\pm1.41) \times
10^{-5},
          \end{eqnarray}
where the uncertainty is conservatively estimated by the size of the strong isospin-breaking.
\begin{table}
                  \centering
                  \renewcommand{\arraystretch}{1.1}
                  \begin{tabular}{c c c  }
                  	\toprule
                  	$Q^2 \ [\mathrm{GeV}^2]$ & $\Delta\alpha_{\mathrm{had}}^{(5)}$ & $(\Delta\sin^2\theta_W)_{\mathrm{had}}$  \\
                  	\noalign{\smallskip}\hline\noalign{\smallskip}
                  	
                  	$0.5$ & 264.0(1.4)(1.4)(1.0)[2.2] & $-268.8(1.5)(1.5)(1.0)[2.3]$  \\
                  	
                  	$1.0$ & 382.9(1.5)(1.1)(1.3)[2.3] & $-390.7(1.6)(1.3)(1.2)[2.4]$  \\ 
                  	
                  	$2.0$ & 516.4(1.7)(1.5)(1.2)[2.6] & $-526.7(1.8)(1.6)(1.2)[2.7]$  \\ 
                  	
                  	$3.0$ & 599.4(1.9)(1.7)(1.3)[2.9] & $-609.8(2.0)(1.8)(1.2)[2.9]$ \\ 
                  	
                  	$4.0$ & 662.1(1.8)(1.4)(1.3)[2.7] & $-672.3(1.8)(1.4)(1.2)[2.6]$   \\
                  	
                  	$5.0$ & 711.2(1.9)(1.4)(1.3)[2.7] & $-720.1(1.8)(1.5)(1.2)[2.6]$  \\
                  	
                  	$6.0$ & 752.3(2.0)(1.5)(1.3)[2.8] & $-759.6(1.9)(1.5)(1.2)[2.7]$ \\ 
                  	
                  	$7.0$ & 787.6(2.1)(1.6)(1.3)[3.0] & $-793.2(2.0)(1.6)(1.2)[2.8]$ \\ 
                  	
                  	$8.0$ & 818.5(2.2)(1.7)(1.3)[3.1] & $-822.4(2.0)(1.7)(1.2)[2.9]$ \\ 
                  	
                  	$9.0$ & 846.4(2.3)(1.8)(1.4)[3.2] & $-848.5(2.1)(1.8)(1.3)[3.0]$  \\ 
                  	
                  	$12.0$ & 916.2(2.4)(2.0)(1.5)[3.5] & $-913.1(2.2)(1.9)(1.3)[3.2]$ \\ 
                  	\bottomrule
                  \end{tabular}
		 \caption{Total HVP contribution to the running of $\Delta\alpha_{\mathrm{had}}^{(5)}$ and
$\sin^2\theta_W$ in isospin-symmetric QCD\@. The first quoted uncertainty is the
statistical error, followed by the systematic error arising from the model
average and the scale setting error.  The final uncertainty in squared
brackets is the sum in quadrature of the previous ones. Results are shown in
units of $ 10^{-5}$. We refer to the companion paper for additional details \cite{companion_paper}.} 
                 \label{tab:dalpha_sin2_space_like_resl} 
  \end{table}
\paragraph{Results}
We compute the contributions to the subtracted HVP at the physical point from the different flavor channels for multiple values of the
squared momentum transfer in the range   $0.25\  \mathrm{GeV}^2\leq Q^2 \leq 12
\ \mathrm{GeV}^2$. Linear combinations of these contributions yield $\bar{\Pi}^{(\gamma,\gamma)}$ and $\bar{\Pi}^{(Z,\gamma)}$, which determine the running of $\alpha$ and $\sin^2\theta_W$ at space-like momenta (see Eq.~(2.16) of Ref.~\cite{companion_paper}). Our calculation includes the effects from $u$, $d$, $s$ and $c$ quarks; the small, missing $b$-quark contribution is estimated using HPQCD determinations of the lowest HVP moments~\cite{Colquhoun:2014ica}. The resulting values for $\Delta\alpha_{\rm had}^{(5)}$ and $(\Delta\sin^2\theta_W)_{\mathrm{had}}$ are listed in Table~\ref{tab:dalpha_sin2_space_like_resl}.              
Figure~\ref{fig:comparison_space_like_region} compares our determination of $\Delta\alpha_{\mathrm{had}}^{(5)}(-Q^2)$ with other  lattice and dispersive evaluations. 
While phenomenological analyses (``DHMZ
data'' \cite{Davier:2019can}, ``Jegerl.\ $\mathtt{alphaQED19}$'' \cite{Jeger_yellow_rep, Jegerlehner:2019alphaQEDc19} and ``KNT18 data'' \cite{Keshavarzi:2018mgv}) are mutually consistent, they lie systematically below our lattice results and those of ``BMWc17'' \cite{Budapest-Marseille-Wuppertal:2017okr}, ``BMWc20'' \cite{Borsanyi:2020mff}. At $Q^2=1 \ \mathrm{GeV}^2$ we observe a discrepancy of up to  $7 \, \sigma$ between
the phenomenological estimates and our value. The tension decreases at higher virtualities but remains sizable, around $4.5\, \sigma$,
at $Q^2 = 9 \ \mathrm{GeV}^2$. 
To provide a continuous and analytically tractable representation of our results, we describe the space-like momentum dependence of the HVP in isospin-symmetric QCD using  Pad\'e approximants, 
\begin{align}
   \bar\Pi(-Q^2) \approx R_M^N(Q^2) = 
          \frac{\sum_{j=0}^{M} a_j Q^{2j} }
         {1 + \sum_{k=1}^{N}b_kQ^{2k}}, 
\end{align}
with the coefficients and correlation matrices  for $\bar\Pi^{(\gamma,\gamma)}$ and  $\bar\Pi^{(Z, \gamma)}$ provided in \cite{companion_paper}. These Pad\'e fits accurately reproduce the lattice data  and are suitable for phenomenological applications.
In order to obtain an estimate at the $Z$-pole, we combine our lattice determinations at space-like momenta with perturbative QCD via the Euclidean split technique \cite{Eidelman:1998vc, Jegerlehner:2008rs}.  The perturbative running is evaluated using the \texttt{AdlerPy} package \cite{Hernandez:2023ipz}, which enables a straightforward inclusion of up-to-date input parameters such as quark masses, together with their associated uncertainty. As an independent  cross-check, we also used the \texttt{pQCDAdler} code, previously employed in \cite{Ce:2022eix}, and updated it to allow for a consistent choice of inputs. With this modification, the \texttt{pQCDAdler} result moves substantially closer to that obtained with \texttt{AdlerPy}, with the residual difference reflecting the different renormalization schemes adopted by the two codes. Our final result for the running at the $Z$-pole, obtained by matching lattice and perturbative QCD at $Q^2=9\ \mathrm{GeV}^2$ and including isospin-breaking corrections, reads
\begin{equation}
\label{res_final}
\Delta\alpha_{\rm had}^{(5)}(M_Z^2) = 0.027821(34)_{\rm lat}(35)_{\rm pQCD},
\end{equation}
which improves upon our previous result~\cite{Ce:2022eix} by a factor of 3, achieving a relative uncertainty of $1.7$\textperthousand.

\begin{figure}
          \centering
          \includegraphics[scale=0.3]{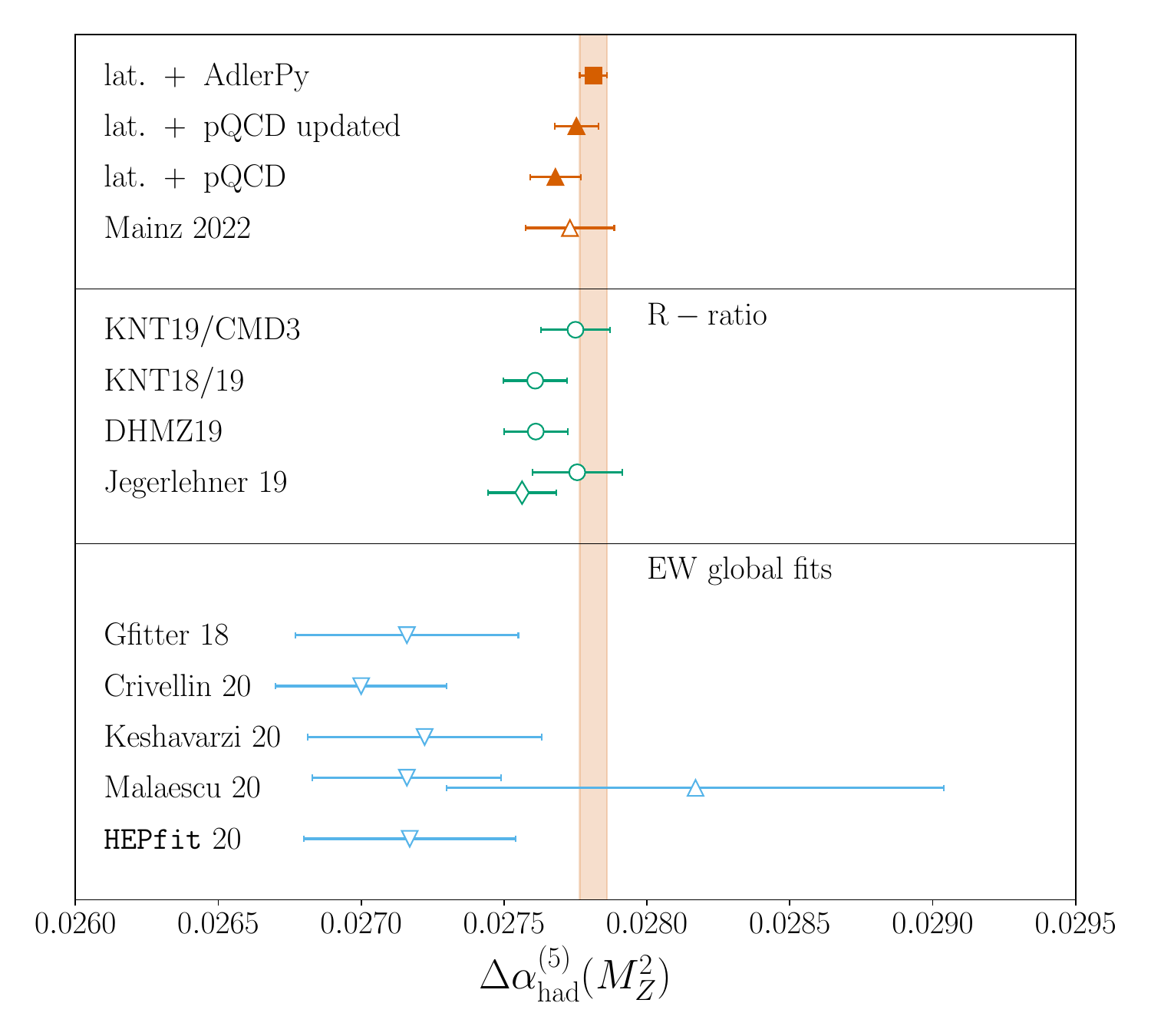}
	 \caption{Summary of results for
$\Delta\alpha_{\mathrm{had}}^{(5)}(M_Z^2)$. The first point shows our
determination using perturbative results from \texttt{AdlerPy} and including isospin-breaking
corrections. The second and third points are obtained using the \texttt{pQCDAdler} software package, followed by the Mainz 2022 result \cite{Ce:2022eix}. Green circles denote results from  standard dispersive approaches based on the $R$ ratio, while blue symbols represent values from global EW fits. See the main text and the companion paper \cite{companion_paper} for additional details.
} 
         \label{fig:z_pole_comparison} 
  \end{figure}

In  Fig.~\ref{fig:z_pole_comparison} we compare our determination in Eq.~(\ref{res_final}) with dispersive approaches and global EW fits. Our result, shown as the filled square, is consistent with our previous  ``Mainz 2022" value \cite{Ce:2022eix}, and the determinations obtained using either \texttt{AdlerPy} or the modified \texttt{pQCDAdler} code are consistent with each other. The dispersive results (green circles) lie below our value at the $1-2\,\sigma$ level, while the determination of \cite{DiLuzio:2024sps} based on CMD-3 data for $e^+e^- \to \pi^+\pi^-$ \cite{CMD-3:2023alj, CMD-3:2023rfe} is fully compatible with our estimate. The tension observed at space-like momenta (cf. Fig.~\ref{fig:comparison_space_like_region}) is less pronounced at the $Z$-pole, due to the additional uncertainty from perturbative running. We also compare to global EW fits (blue open
lower triangles), including results from the  Gfitter group~\cite{Haller:2018nnx},
from~\cite{Crivellin:2020zul} (obtained using the \texttt{HEPfit}
code~\cite{DeBlas:2019ehy}), from~\cite{Keshavarzi:2020bfy, Malaescu:2020zuc}
(obtained from the Gfitter library), and from~\cite{deBlas:2021wap}. These typically favor lower values with reduced precision. The largest deviation, with~\cite{Crivellin:2020zul}, amounts to $2.7\,\sigma$, while all other determinations are compatible within $2\,\sigma$. 

For the electroweak mixing angle, we provide a precise lattice determination of 
the octet-singlet mixing contribution, which is observed on the lattice to approach a constant at large virtualities and therefore tends, in the large momentum limit and in units of $10^{-5}$, to
\begin{equation}
\bar\Pi^{(0,8)}(-Q^2) \xrightarrow{Q^2\to\infty} -\frac{1}{6\sqrt{3}}\Pi^{(0,8)}(0) = 64.0(1.1),
\end{equation}
in good agreement with the phenomenological models given in appendix D of~\cite{Ce:2022eix} but with significantly higher precision.
\paragraph{Prospects for future precise determinations}
While the precision achieved here for $\Delta\alpha_{\mathrm{had}}^{(5)}(M_Z^2)$ already improves phenomenological estimates by a factor of 2, further progress is  required to fully exploit the physics potential of future colliders such as the FCC-ee. In the present  determination, the hadronic running is computed non-perturbatively up to a Euclidean matching scale $Q_0^2$, above which the evolution to the $Z$-pole is obtained from the Adler function in pQCD. These two components exhibit different dependencies on the matching scale. Increasing $Q_0^2$ reduces the interval over which pQCD is applied, therefore reducing its contribution to the uncertainty. At the same time, this requires access to larger values of Euclidean momenta on the lattice, where discretization effects become more prominent. The two error sources thus pull in opposite directions, and the choice of $Q_0^2$ must balance these competing effects.
To assess how future improvements propagate to the final uncertainty at the $Z$-pole, we have constructed a parametric model of the total error  based on the precision of the theory inputs used in pQCD, the lattice uncertainty at the matching point, and the choice of the matching scale itself. On the perturbative side, we parametrize the  uncertainty as
\begin{equation}
	\sigma^2_{\mathrm{pQCD}}(Q_0^2) = \sum_{i,j}^{} g_i(Q_0^2) C_{ij} g_j(Q_0^2) + \sigma^2_{\mathrm{trunc}}(Q_0^2),
\end{equation}
where $g_i$ denotes the sensitivity of $\Delta\alpha_{\mathrm{had}}^{(5)}(M_Z^2)$ to variations of the  theory inputs $p_i\in\{\alpha_s(M_Z), m_c(m_c), m_b(m_b)\}$, namely the strong coupling constant and the charm- and bottom-quark masses, and $C_{ij}$ is their covariance matrix. The second term, $\sigma_{\mathrm{trunc}}^2$, accounts for the truncation uncertainty associated with  the size of the last known  term in the perturbative expansion. On the lattice side, the uncertainty is parametrized as a function of $Q_0^2$ using the error budget of the present calculation at $Q_0^2=9\ \mathrm{GeV}^2$ as a reference. This framework enables a systematic exploration of hypothetical improvement scenarios via multiplicative factors $f_i=\sigma_{p_i} / \sigma_{p_i}^{\mathrm{curr}}$ that rescale the current uncertainties. The total projected variance of  $\Delta\alpha_{\mathrm{had}}^{(5)}(M_Z^2)$ is given by
\begin{equation}
	\sigma_{\mathrm{tot}}^2(Q_0^2; \{f_i, f_{\mathrm{lat}}\}) = \sigma_{\mathrm{pQCD}}^2(Q_0^2; \{f_i\}) +
	\sigma_{\mathrm{lat}}^2(Q_0^2; f_{\mathrm{lat}}).
\end{equation}
For each  scenario, the total projected uncertainty can be evaluated as a function of the matching scale, and the combinations of improvements that are required to reach some target precision can be identified.
\begin{figure}[b!]
	\centering
	\includegraphics[scale=0.38]{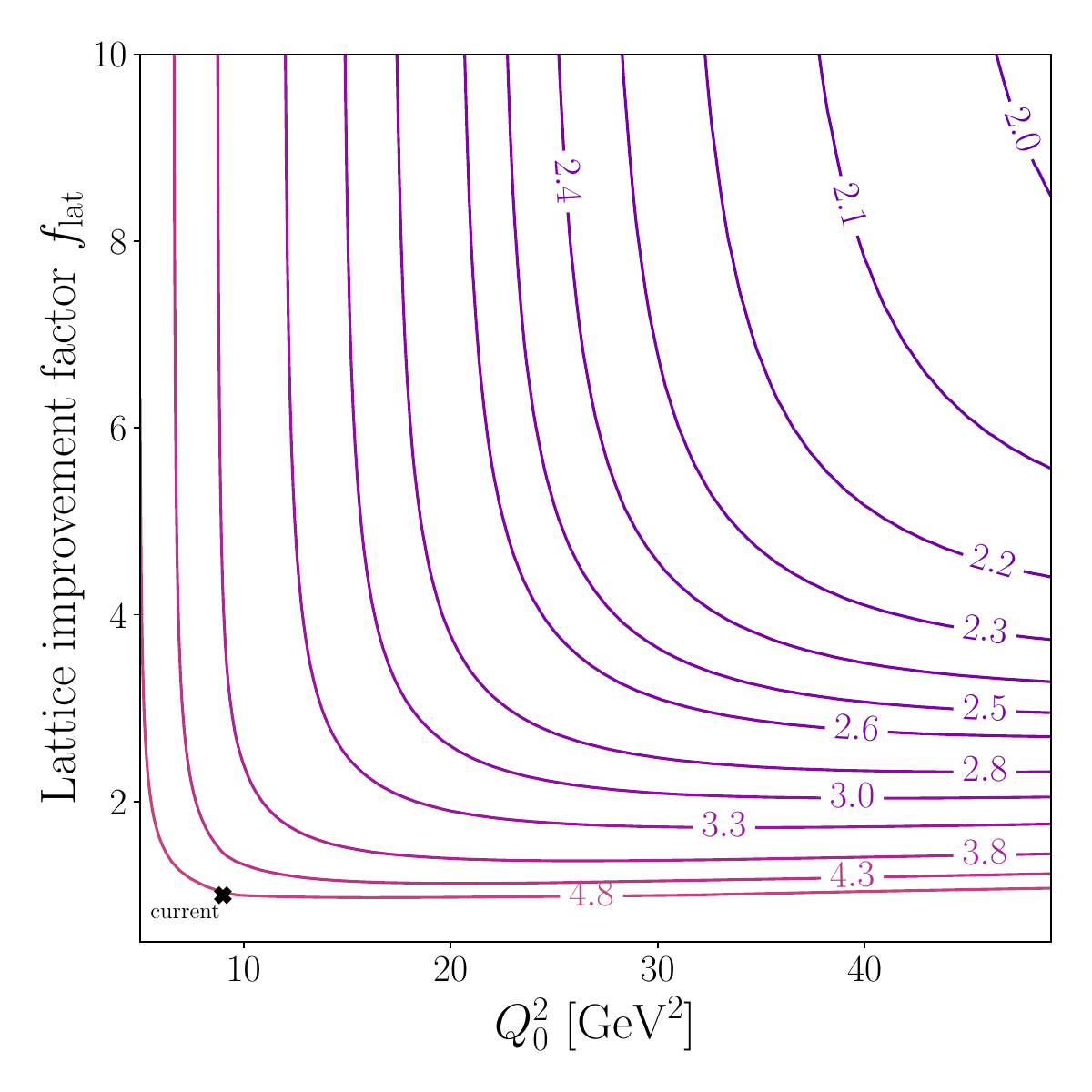}
	\caption{Projected total uncertainty $\sigma_{\mathrm{tot}}$ in the $(Q_0^2, f_{\mathrm{lat}})$ plane. Shown are contour lines of the total error in units of $10^{-5}$, as a function of the matching scale and of a global lattice improvement factor which rescales the current uncertainty.
	} 
	\label{fig:mom_flat_plot} 
\end{figure}
Figure~\ref{fig:mom_flat_plot}  shows contour lines of the projected total uncertainty on $\Delta\alpha_{\mathrm{had}}^{(5)}(M_Z^2)$ in the plane defined by the matching scale $Q_0^2$ and a global lattice improvement factor $f_{\mathrm{lat}}$. No improvement in the pQCD inputs is assumed. Contours are given in units of $10^{-5}$, and the black cross labeled \textquotedblleft current\textquotedblright~denotes  the present determination in Eq.~(\ref{res_final}), obtained at $Q_0^2=9\ \mathrm{GeV}^2$ with $f_{\mathrm{lat}}=1$. Increasing the matching scale alone has little or no effect regarding the reduction of the total uncertainty.  Likewise, significantly more precise lattice calculations at the current matching scale yield only limited gains in $\Delta\alpha_{\mathrm{had}}^{(5)}(M_Z^2)$. In contrast, combining  moderate lattice improvements with a moderately increased matching scale leads to  a substantially more efficient error reduction. In particular, the target precision of $3\times 10^{-5}$ anticipated for FCC-ee measurements \cite{Janot:2015gjr, Freitas:2019bre, Jeger_yellow_rep} can be reached by reducing the lattice uncertainty by $50\%$ or more and increasing the matching scale to $Q_0^2\sim20\ \mathrm{GeV}^2$.

Improvements in the pQCD inputs play a secondary but non-negligible role, with reduced uncertainties in $\alpha_s$ being more relevant than further improvements in the charm and bottom quark masses. The dependence on theory inputs also weakens at larger matching scales, indicating that beyond $Q_0^2\sim 25 \ \mathrm{GeV}^2$ further increases of the matching scale offer diminishing returns.
Finally, reaching the highest precision  foreseen in \cite{Riembau:2025ppc} would require substantial further progress. Our analysis indicates that  factor five improvements in both the lattice determination and the pQCD inputs, together with reduced  perturbative truncation uncertainties, are needed  to match a projected statistical sensitivity of $10^{-5}$ on $\Delta\alpha_{\mathrm{had}}^{(5)}(M_Z^2)$. In this context, lattice QCD can play a central role in reducing the uncertainty of SM parameters, as already demonstrated by state-of-the-art lattice determination of $\alpha_s$ \cite{DallaBrida:2022eua, DallaBrida:2026kuo}.

\paragraph{Conclusion}
We presented a high-precision, first-principles determination of the
hadronic contribution to the running of $\alpha$ and $\sin^2\theta_W$ across a broad range of space-like momenta. Our calculation yields a determination of $\Delta\alpha_{\mathrm{had}}^{(5)}(M_Z^2)$ with a total relative uncertainty of $1.7$\textperthousand, reducing the dominant
non-perturbative uncertainty entering electroweak precision tests. The results exhibit significant tensions with data-driven HVP determinations at space-like virtualities, and provide a robust, first-principles input for future $e^+e^-$ colliders aiming to probe BSM physics. We have performed a quantitative study of future improvement scenarios, analyzing how lattice and pQCD uncertainties, together with the choice of the matching scale, affect the achievable precision on $\Delta\alpha_{\mathrm{had}}^{(5)}(M_Z^2)$. The results show that moderate lattice improvements are most effective to reach the one-permille precision required by future electroweak measurements, while reaching more ambitious targets will additionally require improved pQCD inputs and reduced truncation uncertainties.
Future work will extend the accessible momentum range, incorporate full QED and isospin-breaking  effects, and explore covariant coordinate-space methods to further reduce discretization errors.
\paragraph{Acknowledgments}
We thank Andrew Hanlon, Nolan Miller, Ben H\"{o}rz, Daniel Mohler, Colin Morningstar and Srijit Paul for the collaboration on the data generation and analysis for the spectral reconstruction.
We thank Volodymyr Biloshytskyi for providing auxiliary data used in the 
estimate of isospin-breaking effects.  
A.C. is grateful to Arnau Beltran for valuable discussions and cross-checks performed throughout the analysis.
We are grateful to Marco Cè for sharing data used to produce comparison plots.
We are grateful  to Bogdan Malaescu and the authors of ref. \cite{Davier:2019can} for sharing their data for the running of $\alpha$ at $Q^2>0$.
We thank Alex Keshavarzi and the authors of ref. \cite{Keshavarzi:2018mgv} for sharing the $R$ ratio data with covariance matrix, which were used to calculate tabulated data for the running of $\alpha$ at $Q^2>0$.
We are grateful to our colleagues in the CLS initiative for sharing ensembles.
Calculations for this project were performed on the HPC
clusters Clover and HIMster-II at the Helmholtz Institute Mainz and
Mogon-II and Mogon-NHR at Johannes Gutenberg-Universität (JGU)
Mainz, as well as on the GCS Supercomputers JUQUEEN and JUWELS at 
the Jülich Supercomputing Centre (JSC), HAZELHEN and HAWK at the 
Höchstleistungsrechenzentrum Stuttgart (HLRS), and SuperMUC at the 
Leibniz Supercomputing Centre (LRZ).
The authors gratefully acknowledge the support of the Gauss Centre
for Supercomputing (GCS) and the John von Neumann-Institut für
Computing (NIC) by providing computing time via the projects HMZ21, HMZ23
and HINTSPEC at JSC, as well as projects GCS-HQCD and GCS-MCF300 
at HLRS and LRZ. We also gratefully acknowledge the scientific 
support and HPC resources provided by NHR-SW of Johannes 
Gutenberg-Universität Mainz (project NHR-Gitter).
This work has been supported by Deutsche Forschungsgemeinschaft
(German Research Foundation, DFG) through the Collaborative Research 
Center 1660 “Hadrons and Nuclei as Discovery Tools”, under Grant No. HI~2048/1-2
(Project No.\ 399400745), and through the Cluster of Excellence
``Precision Physics, Fundamental Interactions and Structure of
Matter'' (PRISMA+ EXC 2118/1), funded within the German Excellence
strategy (Project No.\ 390831469). This project has received funding
from the European Union's Horizon Europe research and innovation
programme under the Marie Sk\l{}odowska-Curie grant agreement
No.\ 101106243. 
A.R. was supported by the programme Netzwerke 2021, an initiative of the Ministry of Culture and Science of the State of Northrhine Westphalia, in the NRW-FAIR network, funding code No. NW21-024-A.
\nocite{apsrev41Control}
\bibliographystyle{apsrev4-2}
\bibliography{biblio}

\end{document}